\documentclass[preprint,12pt,3p]{elsarticle}

\usepackage{amssymb}
\usepackage{amsmath}

\begin{document}
\begin{frontmatter}

\title{Nonstandard Null Lagrangians and Gauge Functions and\\ 
Dissipative Forces in Dynamics}

\author{A. L. Segovia, L. C. Vestal and Z. E. Musielak}
\address{Department of Physics, University of Texas 
at Arlington, Arlington, TX 76019, USA \\}

\begin{abstract}
Standard and non-standard Lagrangians that give the same 
equation of motion are significantly different in their forms, as the latter 
do not have terms that clearly discernable energy-like expressions.  A special 
family of these Lagrangians are null Lagrangians and their gauge functions.  
It is shown that non-standard null Lagrangians and their gauge functions 
can be used to introduce dissipative forces to dynamical systems.  With 
standard null Lagrangians being known for introducing non-dissipative 
forces, the presented results allow for a complete picture of novel roles 
played by null Lagrangians in introducing forces to dynamics.  Applications 
of the results to Newton's laws are presented and their Galilean invariance 
is discussed.
\end{abstract}

\end{frontmatter}

\section{Introduction}\label{sec1}

The role of standard Lagrangians (SLs), whose kinetic and potential energy-like terms can 
easily be identified, has been well established in Classical Mechanics (CM) (e.g., [1- 5]) and 
other areas of physics [6].  On the other hand, the so-called non-standard Lagrangians  (NSLs), 
in which neither kinetic nor potential energy-like terms are present, have been introduced to 
CM in recent years (e.g., [7-13]).  The fact that the SLs and NSLs give the same equations 
of motion is well-known.  Moreover, since the original work of Lagrange [1], the physical 
meaning of SLs is also known.  However, the physical meaning of the NSLs remains unclear 
and they are typically treated as mathematical generating functions that accidentally give the 
same equations of motion as the SLs.  To gain more insight into the physical meaning of 
the NSLs is one of the main goals of this Letter.

An additional family of Lagrangians of interest is that of null Lagrangians (NLs), the main 
characteristics of which are as follows: (i) they identically satisfy the Euler-Lagrange (E-L) 
equation, and (ii) they can be expressed as the total derivative of any scalar function, 
called here a gauge function. The properties and applications of these NLs have been 
extensively explored in different fields of mathematics (e.g., [14-20]), and to some 
extend in elasticity [21,22]; however, their physical meaning still remains elusive.  
More recent physical applications of the NLs have involved restoring Galilean invariance 
of the SL in Newtonian dynamics [23,24], and introducing non-dissipative forces to 
CM [25-27], which was done independently from the original Newton approach and 
other proposed approaches [28,29].  It was also shown that there is a broad range 
of forces, and nonlinearities, well-known in classical dynamics that can be accounted 
for by using the NLs [30].  

The forms of SLs and NSLs are distinct, thus, there are also the corresponding 
NLs whose forms resemble those two families of Lagrangians.  In the previous work 
on NLs [23-30], so-called standard NLs have been used.  Recently, non-standard 
NLs have been introduced [31], and this Letter is devoted to their studies. Our main 
aim is to understand the physical meaning of the non-standard NLs and their role in 
dynamical systems.  The results presented in this Letter demonstrate that the 
non-standard NLs can be used to introduce dissipative forces, and that this distinguishes 
them from the standard NLs that are responsible for non-dissipative forces.  From a 
physical point of view, this means that the standard Lagrangians are more suitable 
for describing undamped dynamical systems, while the non-standard Lagrangians are 
more applicable to damped systems; this statement applies equally to the SLs and 
NSLs as well as to the standard and non-standard NLs.  The obtained results are 
used to introduce dissipative forces to the law of inertia and convert it into the 
second Newton law; Galilean invariance of the laws and their Lagrangians is also 
investigated and discussed.

\section{Lagrangians for the law of inertia}\label{sec2}

The SL for the law of inertia is $L_{s}(\dot{x}) = 
\dot {x}^2 / 2$ [2,4], and its substitution into the E-L equation gives 
the equation of motion, $\ddot x  = 0$.  Since the SL
does not depend explicitly on time, the total energy, $E_{\rm tot} 
(\dot x) = \dot {x}^2 / 2$ = const, because $\dot x $ = const, and 
the energy function $E_{s} (\dot x) = L_{s} (\dot x) = E_{\rm tot}
(\dot x)$.  Thus, $dE_{s} / dt = - (\partial L_{s} / \partial t)$ is 
the same as the E-L equation for $L_{s}(\dot{x})$ [4,5]. 

The same equation of motion can be obtained by using the following 
NSL [7-13] 
\begin{equation}
 L_{ns}(\dot{x},x,t)={\frac{1}{g_1(t)\dot{x}+g_2(t) x +g_3(t)}}\ ,
\label{eq1a}
\end{equation}
where $g_{1}(t),g_{2}(t)$, and $g_{3}(t)$ are arbitrary, but at least twice 
differentiable, scalar functions of time $t$.   These functions are determined 
by substituting $ L_{ns}(\dot{x},x,t)$ into the E-L equation and comparing
the form of the resulting equation of motion directly to $\ddot x = 0$.  

Since $ L_{ns}(\dot{x},x,t)$ depends explictly on time, the energy function 
$E_{ns} = \dot x (\partial L_{ns} / \partial \dot x) - L_{ns}$ must be calculated 
[4,5], and the result is 
\begin{equation}
E_{ns}(\dot{x},x,t)= - \left [ 1 + g_1  \dot x  L_{ns}(\dot{x},x,t)
\right ] L_{ns}(\dot{x},x,t)\ ,
\label{eq1b}
\end{equation}
which shows that $E_{ns}(\dot{x},x,t)$ is different than $E_{s} (\dot x)$
obtained above.  Despite the difference, $dE_{ns} / dt = - (\partial L_{ns} 
/ \partial t)$ reduces to the E-L equation when $E_{ns}(\dot{x},x,t)$
and $L_{ns}(\dot{x},x,t)$ are substituted into Eq. (\ref{eq1b}). Thus, as 
expected, the equation of motion resulting from $L_{ns}(\dot{x},x,t)$ is 
identical regardless whether the equation for $E_{ns}(\dot{x},x,t)$ or the
E-L equation is used.

The functions $g_{1}(t),g_{2}(t)$, and $g_{3}(t)$ were evaluated for the law 
of inertia, and two different sets of these functions were found.  For the first 
set: $g_1  = C_1 f^3(t)$, $g_2  = - C_1 a_o f^2(t)$ and  $g_3  = 
C_1 C_2 f^2(t)$, where $C_1$ and $C_2$ are constants of integration, $f(t) 
= a_o t+v_o$, and $v_o$ and $a_o$ are specified by the initial conditions 
required to solve the auxiliary differential equation [31].  For the second set: 
$g_1  = C_3$, $g_2  = 0$ and $g_3 = 0$, where $C_3$ is a constant 
to be determined or simply taken to be $C_3 = 1$ [32].  There are significant 
differences in the forms of these Lagrangians and their energy functions, and 
yet both Lagrangians give the same equation of motion, namely, $\ddot x  
= 0$, which is the law of inertia.

\section{Null Lagrangians and gauge functions}\label{sec3}

The above results show that NSLs can be constructed 
for a given equation of motion.  A method to construct NSLs was 
proposed [31] and it was used to find 
\begin{equation}
L_{n}(\dot{x},x)= \frac{a_1 \dot x}{a_2 x + a_4},
\label{eq2a}
\end{equation}
where $a_1$, $a_2$ and $a_4$ are arbitrary constant coefficients.
It is easy to verify that $L_{n}(\dot{x},x)$ is indeed a null 
Lagrangian, and that its gauge function is  
\begin{equation}
\Phi_{n}(x)= \frac{a_1}{a_2} \ln \vert a_2 x + a_4 \vert\ .
\label{eq2b}
\end{equation}

To generalize these results, it was suggested that the coefficients 
$a_1$, $a_2$ and $a_4$ are replaced by the corresponding functions 
of $t$, so the generalized gauge function [31] becomes
\begin{equation}
\Phi_{gn} (x,t) = {\frac{h_1(t)}{h_2(t)}} \ln \vert h_2(t) x +h_4(t) \vert\ ,
\label{eq3a}
\end{equation}
where $h_1(t)$, $h_2(t)$ and $h_3(t)$ are twice differentiable but 
otherwise arbitrary functions of $t$, and in addition $h_2  \neq 0$.  
This gauge function gives the following general non-standard NL
\[
 L_{gn}(\dot{x},x,t)={\frac{h_1(t)[h_2(t)\dot{x}+{\dot h_2}
(t)x]+{\dot h_4}(t)}{h_2(t)[h_2(t)x+h_4]}}
\]
\begin{equation}
\hskip1.65in +\left[\frac{{\dot h_1}(t)}{h_2(t)}-\frac{h_1(t){\dot h_2}(t)}
{h_2^2(t)}\right] \ln \vert h_2(t)x+h_4  \vert\ ,
\label{eq3b}
\end{equation}
which is significantly different than $L_{n}(\dot{x},x)$ given by 
Eq. (\ref{eq2a}), despite the fact that both are null Lagrangians.

Let $\Phi_{null} [x(t),t]$ be a gauge function (either $\Phi_{n} (x)$ 
or $\Phi_{gn} (x,t)$).  Any NL can then be expressed 
as $L_{null} (\dot x,x,t)  = d\Phi_{null} (x,t) / dt$, which 
gives the energy function 
\begin{equation}
E_{null} (\dot x,x,t) = - \frac{\partial \Phi_{null} (x,t)}{\partial t}\ .
\label{eq4}
\end{equation}
By using $\Phi_{null} (x,t) = \Phi_{n} (x)$, we find $E_{n} = 0$ 
because the gauge function does not depend explicitly on time.  
However, for $\Phi_{null} (x,t) = \Phi_{gn} (x,t)$, we obtain
\[
E_{gn}(\dot{x},x,t)= - \left [ \frac{{\dot h_1}(t)}{h_1(t)}-\frac{
\dot h_2(t)}{h_2(t)} \right] \Phi_{gn}(x,t)
\]
\begin{equation}
\hskip1.65in -\left [\frac{h_1(t)}{h_2(t)}\right] \frac{h_2(t)\dot{x}+
\dot h_2(t)x + {\dot h_4}(t)} {h_2(t)x+h_4(t)}\ .
\label{eq5}
\end{equation}
This shows that different gauge functions have different effects on the 
energy function, namely, for some it can be zero, but for others becomes 
non-zero, and it is interesting that it depends on the gauge function 
itself. 

\section{Introducing forces to the law of intertia}\label{sec4}

Having obtained the relationship between the energy function and 
the gauge function (Eq. \ref{eq4}), and knowing that $E_{null} 
(\dot x,x,t)$ resulting from any non-standard NL is not a NL by 
itself, we may add this extra term to the NSL given by Eq. 
(\ref{eq1a}).  This gives 
\begin{equation}
L_{t}(\dot{x},x,t]= L_{ns}(\dot{x},x,t) - \frac{\partial 
\Phi_{null} (x,t)}{\partial t}\ ,
\label{eq6}
\end{equation}
where the NSL is either 
\begin{equation}
 L_{ns}(\dot{x},x,t)= \frac{1}{C_1 f^2 (t) [f(t) \dot{x}-a_o x+C_2]}\ ,
\label{eq7a}
\end{equation}
or
\begin{equation}
 L_{ns}(\dot{x},x,t)= \frac{1}{C_3 \dot{x}}\ ,
\label{eq7b}
\end{equation}
with $f(t) = a_o t+v_o$, and $\Phi_{null} (x,t) = \Phi_{gn} (x,t)$, 
since there is no contribution from $\Phi_{n} (x)$.

The equation of motion resulting from Eq. (\ref{eq7a}) is 
\begin{equation}
\frac{2 \ddot{x}}{C_1 [f(t) \dot{x} - a_o x + C_2]^3} = 
F (\dot{x},x,t)\ ,
\label{eq8a}
\end{equation}
and using Eq. (\ref{eq7b}), we obtain 
\begin{equation}
\frac{2 \ddot{x}}{C_3 \dot{x}^2} = F (\dot{x},x,t)\ ,
\label{eq8b}
\end{equation}
where the forcing function is 
\begin{equation}
F (\dot{x},x,t) = - \frac{\partial}{\partial x} \left [
\frac{\partial \Phi_{gn} (x,t)}{\partial t} \right ] = 
\frac{\partial E_{gn}(\dot{x},x,t)}{\partial x}\ .
\label{eq8c}
\end{equation}
Because of the presence of additional terms on the LHS of
the above equations, we may write Eqs (\ref{eq8a}) and 
(\ref{eq8b}) in the following forms: $\ddot x = F_1 (\dot x, 
x, t)$ and $\ddot x = F_2 (\dot x, x, t)$, respectively, where
$F_1 (\dot x, x, t) = F (\dot x, x, t) C_1 [f(t) \dot{x} - a_o x 
+ C_2]^3$ and $F_2 (\dot x, x, t) = F (\dot x, x, t) C_3 
\dot {x}^2 / 2$.  The effects of these extra terms on the 
forcing function are discussed in Section 5. 

The above results demonstrate how classical forces can be 
defined using the non-standard NLs, and also show how the 
law of inertia can be converted into the second law of dynamics. 
The main difference between the previous results [25-27] and 
the ones presented in this Letter is the physical nature of the 
forces introduced by the non-standard NLs, namely, the 
forces resulting from the non-standard NLs are dissipative 
(see Section 5), while the forces introduced by the standard 
NLs are non-dissipative.

Our method of converting the first law of dynamics into the 
second one gives an independent way to introduce forces in CM, 
and supplements Newton's definition of forces that directly relates 
them to object's an acceleration and mass [28,29].  Thus, the presented 
results show a deeper connection between Newton's first and second 
laws of dynamics, and demonstrate that non-standard NLs can be used 
to turn undriven dynamical systems into driven ones.  

\section{Physical nature of forces}\label{sec5}

Using Eqs (\ref{eq5}) and (\ref{eq8c}), we obtain the explicit form 
of the forcing function  
\[
F (\dot{x},x,t) = \frac{h_1(t) h_2(t)}{[h_2(t)x(t)+h_4(t)]^2} \left [
\dot x + \left ( \frac{\dot {h}_2(t)}{h_2(t)} - \frac{\dot {h}_1(t)}{h_1(t)}
\right ) x \right ]
\]
\begin{equation}
\hskip1.0in - \frac{h_1(t) h_4(t)}{[h_2(t)x(t)+h_4(t)]^2} 
\left ( \frac{\dot {h}_4(t)}{h_4(t)} - \frac{\dot {h}_1(t)}{h_1(t)} \right )\ .
\label{eq9}
\end{equation}
For $F (\dot{x},x,t)$ to be zero, either $h_1(t) = 0$, nor $h_1(t) = h_2(t) 
= 0$.  There are several special cases, like $h_1(t) = c_1$, $h_2(t) = c_2$ 
and $h_4(t) = 0$, which gives $F (\dot{x},x) = - c_1 \dot {x} / x$, or 
$h_1(t) = c_1$, $h_2(t) = c_2$ and $h_4(t) = c_4$, which results in 
$F (\dot{x},x) = - c_1 c_2 \dot {x} / (c_2 x + c_4)$.  A special case 
of $h_2 (t) = 0$ makes $F (t)$ to be only a function of $t$.  

Other reductions of $F (\dot{x},x,t)$ are also possible but since the forcing
function originates exclusively from the gauge function $\Phi_{gn} (x,t)$,
there are other terms resulting from $L_{ns}(\dot{x},x,t)$ (see Eqs 
\ref{eq8a} and \ref{eq8b}), and these extra terms affect, in addition to
$F_(\dot{x},x,t)$, the forms of functions $F_1 (\dot{x},x,t)$ and 
$F_2 (\dot{x},x,t)$.  Both extra terms depend on the variables $\dot x$,
$x$ and $t$.  Thus, the forces in the equations of motion $\ddot x = F_1 
(\dot x, x, t)$ and $\ddot x = F_2 (\dot x, x, t)$ are always dissipative.  

Thus, based on the presented results, we conclude that the non-standard 
NLs can be used to introduce dissipative forces to dynamical systems.
Since standard NLs are known for introducing non-dissipative forces [25-27], 
our conclusion completes the picture of the novel roles played by the NLs 
in introducing forces to dynamics, and gives new insights into the physical 
meaning of non-standard Lagrangians and non-standard null Lagrangians.

\section{Galilean invariance}\label{sec6}

The problem of Galilean invariance of the law of inertia has been addressed
in most textbooks on CM (e.g., [4,5]).  Some authors also pointed out 
the lack of Galilean invariance of the standard Lagrangian for the law 
of inertia [2,3,23]; however, a solution to this problem that involves 
standard null Lagrangians was recently proposed [24].  An interesting 
question is whether or not the non-standard Lagrangian and the 
resulting forces are Galilean invariant.  Let us now discuss the Galilean 
invariance of the presented results, and refer to all observers who agree
on the Galilean invariance as the {\it Galilean observers}.

The Galilean group of the metric is composed of rotations, translations, 
and boosts between two inertial frames of reference [33,34].   Let $(x,t)$ 
be an inertial frame moving at a constant velocity, $V_0$, with respect 
to a second inertial frame, $(x',t')$, and let their origins coincide at $t=t'
=t_0$. Thus, there are the following transformations between the systems: 
$x'=x-V_{0}t$ and $t'=t$.  By applying these transformations to the law 
of inertia $\ddot x  =0$, it is seen that Newton's first law is Galilean 
invariant.  However, its standard Lagrangian is not [2,4,23] and requires
a special procedure that involves standard null Lagrangians to restore its 
Galilean invariance [24].

Let us now investigate Galilean invariance of the non-standard Lagrangian
for the law of inertia (see Eq. \ref{eq7a}) and written here as 
\begin{equation}
         L_{ns}(\dot{x},x,t)= \frac{1}{C_1 f^2 (t)  [f(t)  \dot{x}
- a_o x + C_2]}\ ,
\label{eq11}
\end{equation}
where $f(t)  = a_o t+v_o$.  After the Galilean transformation, this 
Lagrangian becomes  
\begin{equation}
        L^{\prime}_{ns}[\dot{x}^{\prime}(t^{\prime}),x^{\prime}
(t^{\prime}),t^{\prime}]= \frac{1}{C^{\prime}_1 f^{\prime\ 2} 
(t^{\prime}) [f^{\prime} (t^{\prime}) \dot{x}^{\prime}(t) - 
a^{\prime}_o x^{\prime}(t^{\prime})+C^{\prime}_2 + 
v^{\prime}_o V_0]}\ .
\label{eq12}
\end{equation}
Galilean invariance of $ L_{ns}(\dot{x},x,t)$ requires that its 
form is the same as $L^{\prime}_{ns}[\dot{x}^{\prime}(t^{\prime}),
x^{\prime} (t^{\prime}),t^{\prime}]$.  For the original and transformed
Lagrangians to be of the same form in the variables $x(t)$ and $x^{\prime}
(t^{\prime})$, the following conditions must be satisfied: (i) $f^{\prime} 
(t^{\prime}) = f(t) $, which requires that $a^{\prime}_o = a_o$ and 
$v^{\prime}_o = v_o$; further, it is also required that $t^{\prime} = t$, 
as guaranteed by the Galilean transformation; (ii) $C^{\prime}_1 = C_1$ 
is satisfied in all intertial frames; and (iii) $C^{\prime}_2 + v_o V_0 = C_2$
to be valid for all Galilean observers.

Since $a_o$ and $v_o$ are the integration constants for the auxiliary equation
[31], and $C_1$ and $C_2$ are the constants of integration for the law of inertia,
these constants are determined by the intitial conditions to be specified for a 
physical problem to be solved.  However, both the auxiliary equation and the 
law of inertia are Galilean invariant; thus, the solutions to these equations must 
also be the same (Galilean invariant) for all Galilean observers.  The latter is 
equivalent to the requirement that the specified initial conditions are also the 
same for all Galilean observers, which validates the above conditions (i) and (ii).
The condition (iii) shows that $C^{\prime}_2 \neq C_2$ and that the constant 
$C^{\prime}_2$ must be modified by adding another constant $v_o V_0$ to 
it as compared to $C_2$.  This addition is known in advance by all Galilean 
observers, who by their definition already agreed on the Galilean invariance.
Therefore, the non-standard Lagrangian for the law of inertia given by Eq. 
(\ref{eq7a}) is Galilean invariant, which makes it distinguished from the 
standard Lagrangian, whose original form is not Galilean invariant [2].

Having demonstrated the Galilean invariance of the law of inertia and its
non-standard Lagrangian, we now check the Galilean invariance of the 
equation of motions $\ddot x = F_1 (\dot x, x, t)$ and $\ddot x = F_2 
(\dot x, x, t)$.  It is easy to verify that neither $F_1 (\dot x, x, t)$ nor 
$F_2 (\dot x, x, t)$ is Galilean invariant because they are dissipative 
forces that depend explicitly on both $\dot x$ and $x$.  Thus, the 
Galilean invariance is lost when the law of inertia is converted into 
the second law of dynamics.  

The results of this Letter demonstrate that dissipative forces in CM can 
also be defined using non-standard null Lagrangians, which is a novel 
way to view forces and it significantly extends the previous work on 
standard null Lagrangians that were used to introduce non-dissipative 
forces to CM [25-27].  The presented results also show that the 
non-standard Lagrangian for the law of inertia preserves its Galilean 
invariance, which makes it different from the standard Lagrangian [2,4], 
whose Galilean invariance must be fixed by using a special procedure 
that involves standard null Lagrangians [24].  The presented results give novel insight into the role played by non-standard Lagrangians and
non-standard null Lagrangians, which seem to be more suitable for describing
damped dynamical systems, while standard Lagrangians and standard null 
Lagrangians seem to be more applicable to undamped dynamical systems.

Finally, let us point out that the results presented in this Letter were
obtained within the framework of CM and that they can be extended to 
stochastic processes and their dissipative stochastic description [35,36] 
as well as to classical [37] and quantum [38] fields; however, this will
be addressed elsewehere.

\bigskip\noindent
{\bf Acknowledgments}

We appreciate very much valuable comments brought to our attention 
by an anonymous referee, which allowed us to significantly improve this Letter.  
A.L.S. is grateful to the McNair Scholars Program for the Summer funding of 
this research at the University of Texas at Arlington. 

\end{document}